\documentclass[journal, lettersize]{IEEEtran}
\usepackage{amsmath,amsfonts,amssymb,amsthm,mathtools,bm}
\usepackage{algorithmic}
\usepackage{algorithm}
\usepackage{array}
\usepackage{tabularx}
\usepackage{subcaption}
\usepackage{textcomp}
\usepackage{stfloats}
\usepackage{url}
\usepackage{verbatim}
\usepackage{comment}
\usepackage{graphicx}
\usepackage{subcaption}
\usepackage{cite}
\usepackage{pifont}
\usepackage{multirow}

\usepackage{microtype}
\usepackage{booktabs} 
\usepackage{tikz}
\usepackage[hidelinks]{hyperref}

\usepackage{comment}

\usepackage{listings}
\begin{document}

\title{Deadline-Bound Finite-Object Delivery over Intermittent LEO Satellite Contact Plans under Residual-Service Accounting}
\author{Houtianfu Wang,~\IEEEmembership{Student Member,~IEEE,} O. Tansel Baydas,~\IEEEmembership{Student Member,~IEEE,} Hanlin Cai,~\IEEEmembership{Student Member,~IEEE,} Haofan Dong,~\IEEEmembership{Student Member,~IEEE,} Ozgur B. Akan,~\IEEEmembership{Fellow,~IEEE}
\thanks{Authors are with the Center for neXt Communications (CXC) Group, Electrical Engineering Division, Department of Engineering, University of Cambridge, UK. E-mails:\{hw680, otb26, hc663, hd489, oba21\}@cam.ac.uk 
}
\thanks{Ozgur~B.~Akan is also with the Centre for neXt Communications (CXC), Department of Electrical and Electronics Engineering, Ko\c{c} University, 34450 Istanbul, T\"{u}rkiye. E-mail: akan@ku.edu.tr}}


\maketitle

\begin{abstract}
Low-Earth-orbit (LEO) relay networks deliver finite objects—sensing tiles, telemetry blocks, model updates, and checkpoints—over intermittent inter-satellite and space-to-ground contact plans. Partial delivery is insufficient when the complete object misses its deadline. When an object is split across candidate paths, a path-private evaluation can count the same contact service more than once and silently under-count completion. We develop a residual-service-aware delivery layer that consumes candidate paths from contact-plan route generation and tests whether the complete object can be delivered before its deadline under per-edge first-in-first-out residual service. Under controlled shared-contact contention, path-private evaluation under-counts completion by up to 154 s and can report finite completion for a fixed plan with no residual-service completion. For edge-disjoint complementary contacts, the layer reduces to fixed-path service; we derive a sufficient service-budget condition under which two-way striping strictly enlarges the feasible payload region. We verify a restricted exhaustive reference, characterize runtime over a 20–180-satellite procedural contact model, and show that bounded two-way striping reduces mean and median gaps to the restricted reference by about 40\%, while P90 and worst-case gaps remain unchanged.

\end{abstract}


\section{Introduction}
\label{sec:introduction}

Low-Earth-orbit (LEO) systems increasingly support on-board sensing and processing, producing finite data objects whose value depends on timely delivery to the ground. LEO relay networks that carry these objects can be modeled as contact-plan delay-/disruption-tolerant networks, where inter-satellite and space-to-ground transmission opportunities appear only during scheduled contact windows~\cite{fraire2021routing,cao2022overview}. This paper studies the service layer that decides whether a finite data object can be delivered over such intermittent contacts before its deadline. The object may be a sensing tile, telemetry block, model update, checkpoint object, or control-related data product; the common feature is that the complete object is useful only if it reaches the ground before its deadline.

A key difficulty is that splitting an object across contacts can create a quiet accounting failure. If two chunks of the same object, or competing objects in the same contact window, consume the same contact service, a path-private evaluation can count that service more than once. Such an evaluation can under-count completion time and, under a payload-stress diagnostic, report a finite completion for a fixed plan for which no residual-service completion exists. This failure is not visible from average rate, bottleneck throughput, or link-budget snapshots alone.

The relay infrastructure is itself mobile: LEO nodes move fast enough that connectivity exists only during scheduled contact windows, the RF/optical service rate varies over each contact, and contact capacity is a shared wireless resource. These are the intermittent-connectivity and shared-resource constraints of mobile/non-terrestrial service deployment: intermittent wireless service, shared contact capacity that cannot be double-counted, and hard object deadlines that determine whether a delivered object remains useful. Model updates and checkpoints are examples of finite-object workloads within a communication-side deadline-delivery problem.



For such objects, the relevant communication question is how much complete payload can be delivered inside the release-time--deadline window. A path with a high nominal rate may be unavailable inside the deadline window, while a lower-rate path may complete earlier if its useful service appears earlier after release.

\begin{figure*}[t]
\centering
\includegraphics[width=0.94\textwidth]{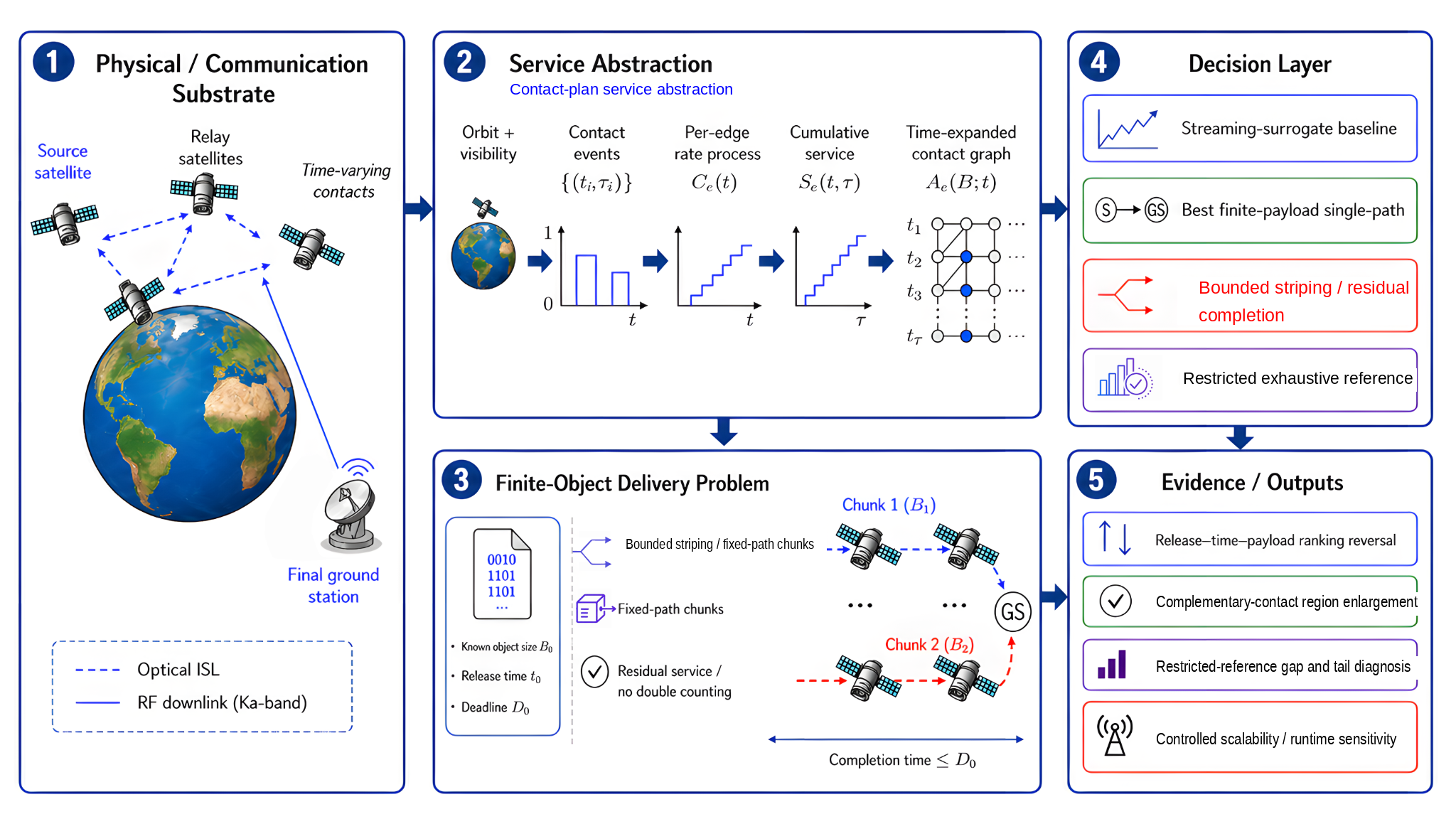}
\caption{Residual-service-aware finite-object delivery over intermittent LEO contact plans.}
\label{fig:framework}
\vspace{-6pt}
\end{figure*}

We therefore develop a plan-conditioned residual-service evaluator that treats the logical object as the completion unit and evaluates each chunk under the service state induced by the candidate plan. For a committed plan, the evaluator tracks per-edge FIFO reservations and keeps contact-service consumption consistent
across transmissions. For edge-disjoint complementary contacts, no per-edge reservation conflict exists; residual evaluation reduces to fixed-path service, and feasible-region gain comes from complementary contact timing. The evaluator supports single-path delivery, bounded two-way delivery decisions, and comparison with a restricted exhaustive reference. Figure~\ref{fig:framework} summarizes the resulting service-layer workflow.

\subsection{Contributions}

This paper is organized around a finite-object delivery service layer for intermittent LEO contact plans. The main contributions are as follows:
\begin{enumerate}
    \item Finite-object completion service layer.
    We formulate deadline-aware finite-object delivery as a service layer placed after contact-plan route generation. Given candidate paths, chunk sizes, and launch times, the layer maps per-edge cumulative RF/optical service to object-level deadline feasibility under per-edge first-in-first-out (FIFO) service.

    \item Residual correctness under shared-contact contention.
    We develop a plan-conditioned residual-service evaluator that prevents shared contact service from being double-counted. The evaluator is non-trivial when chunks of the same object or competing objects share a contact. In controlled contention diagnostics, a path-private evaluation under-counts completion when transmissions share a contact, whether they are chunks of the same object or competing objects; under a payload-stress diagnostic, it can report a finite completion for a fixed plan for which no residual-service completion is found.

    \item Edge-disjoint complementary-contact striping gain.
    We derive a sufficient deadline service-budget condition under which two edge-disjoint path profiles strictly enlarge the candidate-family deadline-feasible payload region after chunk overhead. This gain is a complementary-contact effect, evaluated over a release distribution rather than a single selected release.

    \item Verified restricted reference, tail diagnosis, and scalability characterization.
    We use a restricted exhaustive reference that is exact over its discretized two-way plan family and verify its implementation against un-pruned enumeration. We quantify how much of this restricted-reference benefit is recovered by bounded search. We show that the unrecovered upper tail is associated with single-path-versus-two-way plan-selection disagreement rather than search-grid resolution, and provide controlled scalability characterization: tractable light-arm scaling over 20--180 satellites with a joint-scheduler greedy-fallback runtime boundary.
\end{enumerate}

The remainder of the paper is organized as follows. Sec. II reviews related work. Sec. III introduces the service-layer interface and shared-service model. Secs. IV--VI define the finite-object delivery problem, cumulative-service link model, and multi-hop earliest-arrival evaluator. Sec. VII presents the complementary-contact striping analysis and the restricted exhaustive reference. Sec. VIII gives the controlled evaluation, including residual-accounting correctness, release-distribution results, restricted-reference gap diagnosis, and scalability sensitivity. Sec. IX concludes.

\section{Related Work}
\label{sec:literature}

We review contact-aware routing, multipath and fragmentation, satellite-assisted mobile/edge systems, and communication-service models for finite-object completion over intermittent LEO contact opportunities.

\subsection{Contact-Aware Routing and Contact-Plan Delivery}
\label{sec:lit_routing}

Routing in time-varying LEO satellite networks has been studied from the perspectives of path construction, delay reduction, load balancing, and disruption tolerance. Low-latency routing in space highlights the importance of topology dynamics and propagation delay~\cite{handley2018delay}, while stochastic-geometry and reliability-oriented studies characterize large-scale LEO routing behavior under dynamic connectivity~\cite{wang2022sglatency,wang2024urll}. Recent mobile-computing work has also examined LEO routing and
control-plane support, including software-defined multicast with segment routing, satellite-ground interconnection design,
satellite-terrestrial cooperative routing, source routing, and resilient
multi-attribute routing in mega-constellations~\cite{hu2024sdm,liu2025interconnection,feng2024cooperativeRouting,zhang2025sourceRouting,li2025differentialRouting,li2025mcsrRouting}. These works optimize route
selection, control-plane operation, or network-level delivery performance.


Contact-plan and disruption-tolerant networking provide the route- and contact-planning foundation for our setting. CGR and its variants exploit predicted future contacts in delay-tolerant space networks~\cite{fraire2021routing}, while Schedule-Aware Bundle Routing (SABR) standardizes schedule-aware forwarding for DTN bundles over planned contacts~\cite{ccsds2019sabr}; broader surveys summarize the dynamic routing design space in satellite networks~\cite{cao2022overview}. CGR-ETO and overbooking management incorporate queue-aware contact availability and contact-volume consumption at the bundle-forwarding layer. The present paper uses this contact-consumption view at the finite-object completion layer: candidate chunks, splits, and launch times define a plan-conditioned service state, and the output is whether the complete object meets its deadline rather than the transmission time of an individual bundle. The transmissions that consume edge service are induced by the candidate plan itself, so the per-edge FIFO order and residual service are functions of the plan rather than of an externally fixed backlog. Contact-plan route generators can therefore supply the candidate path family, while the proposed layer evaluates finite-object completion under residual contact service.

\subsection{Multipath, Fragmentation, and Object-Level Deadline Delivery}
\label{sec:lit_multipath}

Multipath cooperative routing has been studied in LEO satellite networks by distributing traffic over multiple paths~\cite{tang2019multipath}. At a different layer, multipath deadline-aware transport mechanisms have been developed for space networks with lossy, long-delay relay links~\cite{shi2021mpdtp}, and fragment-level forwarding has been considered in LEO networks, for example through low-delay fragment forwarding based on named data networking~\cite{diao2023lowdelay}. These works motivate the use of multiple paths, transport-level deadline control, and fragment-level forwarding in space networks.

The present paper places these mechanisms inside a cumulative-service finite-object model. The question is not only whether multiple paths or fragments can be used, but whether a split object completes before its deadline after shared contact service, per-chunk overhead, and residual service consumption are accounted for. This distinction is important because an edge-disjoint two-way split gives complementary-contact gain, whereas shared-edge or cross-object contention requires residual-service accounting to avoid double-counting.

\subsection{Satellite-Assisted Mobile/Edge Systems and Finite-Object Workloads}
\label{sec:lit_mobile_edge}

Recent work places LEO and non-terrestrial networks within the mobile-computing scope through explicit system components, transport support, offloading, and learning workloads. At the transport layer, PEPesc designs a TCP performance-enhancing proxy for
non-terrestrial networks~\cite{li2024pepesc}. For orbital edge computing and
Earth-observation workloads, FOOL addresses the satellite downlink bottleneck by
compressing neural features before transmission~\cite{furutanpey2025fool}. In
satellite-assisted mobile/edge computing, studies have considered ITS data
offloading, peer offloading, service chaining, and resource allocation over
satellite access or satellite-terrestrial edge systems~\cite{hassan2024satelliteITS,zhang2024peerOffloading,xie2025offloadingDelay,chen2025uavLeoOffloading,chen2025gameOffloading,lan2025securityOffloading,zhou2025latencyEnergyOffloading,xia2024serviceChaining}. Learning-oriented systems
over LEO or space-ground networks~\cite{lin2025fedsn,zhai2024fedleo,yang2024progressiveFL,zhu2025hierarchical}
further motivate finite objects such as model updates, checkpoints, and intermediate data products.

These works reduce transport cost, allocate computation, compress
features, or optimize learning workflows. This paper studies the complementary
completion layer: once the finite object size is determined, the layer evaluates
whether intermittent contact service can deliver the complete object before its
deadline and whether split/launch decisions double-count shared contact service.

\subsection{Communication-Service Models for LEO Networking}
\label{sec:lit_model}

A related strand studies LEO networking through communication-model and service-level abstractions rather than through routing-policy design. Downlink and constellation-level analyses characterize coverage, spatial service, and time-varying link availability~\cite{park2023tractable,okati2022nonhomogeneous}. RF/optical satellite links have also been analyzed in terms of latency, outage, power, and service availability~\cite{liang2024free,wang2024serviceblockage}. These models characterize the time-varying communication service supplied by satellite links.

We use such RF/optical service profiles as the input layer and study the next step: how intermittent service, contact timing, and residual contact consumption determine finite-object completion and deadline feasibility. The proposed evaluator is therefore route-generator agnostic and service-profile driven: it consumes candidate paths and per-edge service profiles, then tests object-level completion under residual contact reservation.

\section{System Model and Service-Layer Interface}
\label{sec:system_model}

This section defines the service-layer model used throughout the paper. The layer consumes a contact plan, per-edge communication-service profiles, and a finite candidate path family, and returns object-level completion or deadline-feasibility decisions under a fixed per-edge service discipline. The model is relay-only: intermediate satellites store and forward object chunks but do not execute compute tasks or make placement decisions. The main analysis is single-object finite delivery; controlled contention diagnostics later introduce background objects only to exercise shared-contact residual accounting.

\subsection{Service-Layer Interface and Scope}

The evaluator sits between contact-plan route generation and the transport or application layer. Its inputs are the contact plan, per-edge RF/optical service profiles, a finite candidate path family, and an object descriptor containing size, release time, and deadline. Its outputs are a completion or feasibility decision, together with the selected path assignment, chunk sizes, and launch times for the evaluated plan.

The layer is route-generator agnostic. Candidate paths may come from contact-plan routing, CGR-/SABR-style generation, or the bounded internal path enumerator used for controlled evaluation. The evaluator consumes route-generation outputs and tests whether the candidate service profiles provide enough residual cumulative service for complete-object delivery before the deadline.

\subsection{Object Model}

We first define the bounded finite-object plan family used in the analytical results and in the restricted exhaustive reference. A finite object is delivered from a source satellite to a final ground station and is represented by
\begin{equation}
o=(t_o^{\mathrm{rel}},B_o,\Delta_o),
\end{equation}
where \(t_o^{\mathrm{rel}}\) is the release time, \(B_o\) is the object size in bits, and \(\Delta_o\) is the relative deadline budget. The absolute deadline is
\begin{equation}
d_o=t_o^{\mathrm{rel}}+\Delta_o .
\end{equation}

In the bounded plan family, the object may be transmitted as a whole or partitioned into fixed-path chunks,
\begin{equation}
\sum_{k=1}^{K_o} b_{o,k} = B_o, \qquad K_o \le K_{\max},
\end{equation}
where \(K_o\) is the number of chunks in the bounded plan, \(b_{o,k}\) is the size of chunk \(k\), and \(K_{\max}\) is the striping-degree limit. The analytical two-way condition and the restricted exhaustive reference use \(K_{\max}=2\), which keeps the plan family interpretable and the exhaustive comparison tractable.

The scalable scheduler used in the system experiments additionally includes a quantized greedy fallback outside this restricted \(K=2\) reference family. This implementation behavior is reported separately in the evaluation: it can occasionally complete earlier than the restricted reference, but it is not part of the restricted-reference guarantee. Larger restricted striping degrees and full multi-object scheduling are left for future work.

\subsection{Communication Network Model}

The space--ground network is represented as a time-varying directed graph
\begin{equation}
G(t) = (V, E(t)), \qquad V = V_{\mathrm{sat}} \cup V_{\mathrm{GS}},
\end{equation}
where \(V_{\mathrm{sat}}\) is the set of satellites and \(V_{\mathrm{GS}}\) is the set of ground stations. Topology changes, contact-window boundaries, and channel-state changes are represented by a finite ordered event set
\begin{equation}
0 = \tau_0 < \tau_1 < \cdots < \tau_H .
\label{eq:event_times}
\end{equation}
On each interval \([\tau_\ell,\tau_{\ell+1})\), the active edge set and link-service profile are evaluated from the contact and channel model. The event set is a computational abstraction; it does not imply real ephemeris validation or a specific propagator.

The controlled numerical evaluation uses a procedural Walker-like relay topology with a fixed source satellite and a final ground station. Satellites are connected by optical inter-satellite links to neighbouring satellites, and visible source-to-ground or relay-to-ground edges use RF downlinks. This topology is used as a controlled contact-plan model rather than a TLE/SGP4 ephemeris trace~\cite{vallado2006spacetrack}.

\subsection{Service Discipline and Shared Service Interface}

The network operates under a fixed per-edge FIFO service discipline. Chunk transmission on a selected path is non-preemptive: once a chunk is launched on a fixed path, it is not switched to another path in transit. All chunks of the same object share the same final destination. In controlled contention diagnostics, additional background objects may be inserted into the same contact window to expose shared-contact service consumption; the main analytical benchmark remains single-object.

Because multiple transmissions may request the same contact opportunity, shared edge service cannot be double-counted. For a candidate delivery plan \(\chi\), let \(\mathcal{J}_e(\chi)\) denote the set of transmissions in the plan that traverse edge \(e\). These transmissions are ordered by their arrival times to \(e\) under FIFO service; deterministic tie-breaking is used when arrival times are equal. This order is denoted by \(\prec_e\).

The usable service rate on edge \(e\) is \(C_e(t)\zeta_e(t)\), where \(C_e(t)\) is the instantaneous capacity and \(\zeta_e(t)\) is the edge-availability indicator introduced in Sec.~V. Its integral measures the payload service supplied by that edge. For a transmission \(j\in\mathcal{J}_e(\chi)\), let \(\mathcal{I}^{<j}_e(\chi)\) be the set of service intervals on edge \(e\) already reserved by transmissions \(r\prec_e j\). The residual cumulative service available to \(j\) over \([t,t+\tau]\) is
\begin{equation}
S^{\mathrm{res}}_{e,j}(t,\tau\mid\chi)
=
\int_t^{t+\tau}
C_e(u)\zeta_e(u)
\mathbf{1}
\left\{
u\notin
\bigcup_{I\in\mathcal{I}^{<j}_e(\chi)} I
\right\}
\,du .
\label{eq:residual_service}
\end{equation}
When the evaluated transmission is clear from context, we suppress \(j\) and write \(S^{\mathrm{res}}_e(t,\tau|\chi)\).

This residual service is plan-conditioned: the consuming transmissions are induced by the candidate plan through path choices, chunk sizes, launch times, and FIFO ordering. It prevents contact service from being counted more than once when same-object chunks or controlled competing objects share an edge.

\subsection{Plan-Conditioned Residual-Service Accounting}
\label{subsec:plan_endogenous_accounting}

The residual-service interface distinguishes finite-object completion from ordinary path or bundle evaluation. A route- or bundle-level metric evaluates a delivery opportunity against a service profile that is fixed before the route or bundle is scored, possibly after accounting for exogenous backlog. In contrast, once a finite object is split into chunks, the transmissions that consume service on an edge are induced by the candidate plan itself. Thus, \(\mathcal{J}_e(\chi)\), the FIFO order \(\prec_e\), and the residual service available to each chunk are functions of \(\chi\).

A minimal shared-edge example illustrates the issue. Consider two transmissions \(j_1\) and \(j_2\) that traverse edge \(e\), with \(j_1 \prec_e j_2\), and let \(t_2\) be the time at which \(j_2\) reaches \(e\). A path-private evaluation of \(j_2\) tests completion over the full service \(S_e(t_2,\tau)\). Under plan \(\chi\), the correct test instead uses
\begin{equation}
    S^{\mathrm{res}}_{e,j_2}(t_2,\tau \mid \chi) \ge b_{j_2},
\end{equation}
where service already reserved by earlier transmissions is removed. Since
\begin{equation}
    S^{\mathrm{res}}_{e,j_2}(t_2,\tau \mid \chi) \le S_e(t_2,\tau),
\end{equation}
with strict inequality whenever the evaluation window overlaps reserved service with positive rate, path-private evaluation can report a finite completion even when no residual-service window before the deadline can carry the chunk. This is the failure mode evaluated under same-object and cross-object contention in Sec.~VIII-A.

This comparison is a fixed-plan statement: it compares path-private and residual evaluations of the same candidate plan \(\chi\). It does not imply that a residual-evaluated scheduler is bounded above or below by the restricted exhaustive reference, because the scheduler and the reference may commit different plans and may search different plan families. In particular, the scalable scheduler's greedy fallback can lie outside the restricted \(K=2\) reference family, as reported in Sec.~VIII.

When candidate chunk paths are edge-disjoint, no per-edge reservation conflict occurs within the plan, and the residual evaluation reduces to unconditioned fixed-path service. This is the regime used in the complementary-contact sufficient condition of Proposition~2. Thus, residual accounting and complementary-contact striping gain are distinct effects: the former is a correctness requirement under shared-contact contention, while the latter is a timing gain from edge-disjoint service opportunities.

The residual-service accounting layer can be used after any candidate path-generation step, including CGR- or SABR-style route generation, because it operates on a finite candidate path family rather than on the route-generation procedure itself.





\subsection{Object-Level Communication Primitive}

We use \(A_{m\to n}(B;t)\) to denote the communication delay
from releasing a finite payload of \(B\) bits at node \(m\) at
time \(t\) until the complete payload has arrived at node \(n\).
In the notation below, \(A_e\) denotes the single-edge completion
delay, \(A_P\) denotes the completion delay along a fixed path,
and \(A_{m\to n}\) denotes the best finite-payload earliest-arrival
delay over the generated path family. The single-edge and
fixed-path versions are constructed in Secs.~V and VI, and the
generated-family best-path version is defined in Sec.~VI-C. A
vertical bar ``\(|\chi\)'' indicates evaluation under the residual
service state induced by a candidate delivery plan. Thus,
\(\chi\)-conditioned quantities are used for fixed-plan residual
evaluation, whereas unconditioned best-path quantities are used
for candidate generation and same-model non-striping baseline
evaluation.

\section{Finite-Object Delivery Problem Formulation}
\label{sec:problem_formulation}

Building on the residual-service accounting layer introduced in
Sec.~\ref{subsec:plan_endogenous_accounting}, this section formalizes the
communication-centric finite-object delivery problem studied in the paper.
The formulation focuses on relay-only delivery of a finite object from a
source satellite to a final ground station over time-varying contact
opportunities.

\subsection{Delivery Plan and Chunk-Level Variables}

Within the bounded plan family, a candidate delivery plan for object \(o\) is denoted by
\begin{equation}
\chi = \left\{ \left( b_{o,k}, t_{o,k}^{\mathrm{tx}}, P_{o,k} \right) \right\}_{k=1}^{K_o},
\end{equation}
where $b_{o,k}$ is the size of chunk $k$, $t^{\mathrm{tx}}_{o,k}$ is its launch time, and $P_{o,k}$ is a fixed time-respecting path from the source satellite $s_o$ to the final ground station $g_o$. The launch time of every chunk must satisfy
\begin{equation}
t^{\mathrm{tx}}_{o,k}\ge t_o^{\mathrm{rel}},\quad \forall k.
\end{equation}

Each path $P_{o,k}$ is selected from the feasible path set induced by the time-varying contact network. Once launched, chunk $k$ is transmitted non-preemptively along $P_{o,k}$ and cannot switch to another path in transit. All chunks of the same object share the same final ground-station destination.

Each chunk incurs a per-chunk overhead modeled in equivalent time units. We write
\begin{equation}
\eta_{\mathrm{chunk}} = \eta_{\mathrm{hdr}} + \eta_{\mathrm{setup}},
\end{equation}
where $\eta_{\mathrm{hdr}}$ denotes an equivalent time overhead
associated with lightweight header or bookkeeping operations,
and $\eta_{\mathrm{setup}}$ denotes an equivalent time overhead
associated with chunk-level setup, acquisition, or switching
penalties. Any bit-level protocol overhead is absorbed into
these equivalent time quantities under the same communication
service model.

\subsection{Chunk-Level Arrival and Object Completion Time}

Given a candidate delivery plan $\chi$, the arrival time of chunk
$k$ is defined along its selected fixed path $P_{o,k}$ as
\begin{equation}
T^{\mathrm{arr}}_{o,k}(\chi)
=
t^{\mathrm{tx}}_{o,k}
+
A_{P_{o,k}}
\bigl(
b_{o,k};t^{\mathrm{tx}}_{o,k}\mid \chi
\bigr),
\end{equation}
where $A_{P_{o,k}}(\cdot \mid \chi)$ denotes the $\chi$-conditioned
finite-payload delay (defined in Sec.~VI) along the fixed path $P_{o,k}$
selected in the delivery plan. Thus, once $\chi$ specifies $P_{o,k}$, the chunk is evaluated on that fixed path and is not re-routed by another minimization over paths. The best-path evaluator $A_{s_o\to g_o}(\cdot)$ is used only when constructing candidate paths or when evaluating the non-striping
single-path baseline.

The object completion time is then
\begin{equation}
T_o^{\mathrm{cmp}}(\chi)
=
\max_{1 \le k \le K_o} T_{o,k}^{\mathrm{arr}}(\chi)
+
T_o^{\mathrm{reasm}}
+
K_o \eta_{\mathrm{chunk}},
\end{equation}
where $T_o^{\mathrm{reasm}}$ denotes a final reassembly or completion overhead at the destination.

The completion metric is object-centric: a finite object
completes only when its last chunk arrives and all chunk-level
overheads have been absorbed. Since \(T_o^{\mathrm{cmp}}(\chi)\)
is an absolute time, deadline feasibility is evaluated through
the release-relative condition
\begin{equation}
T_o^{\mathrm{cmp}}(\chi)-t_o^{\mathrm{rel}}\le \Delta_o.
\end{equation}

\subsection{Residual-Service-Consistent Feasibility}

The delivery plan \(\chi\) is feasible within the bounded plan family used by the analytical results and the restricted exhaustive reference only if it respects both object conservation and shared service constraints. The implemented scalable scheduler may additionally invoke a quantized greedy fallback outside this restricted family. First, the chunk sizes must satisfy
\begin{equation}
\sum_{k=1}^{K_o} b_{o,k} = B_o,
\qquad
K_o \le K_{\max},
\qquad
b_{o,k} > 0.
\end{equation}

Second, each selected path must be time-respecting and compatible with the event-driven contact graph. Third, shared edge service must be accounted for consistently through the residual service profiles induced by $\chi$. This means that the completion of each chunk is evaluated under the service state left by previously scheduled transmissions and that the same contact service cannot be assigned twice.

Thus, chunk-level evaluations are coupled through shared service accounting
rather than independent empty-network path computations.

\subsection{Bounded Finite-Object Completion Problem}

On top of the above definitions, we use a bounded finite-object completion problem to define the reference optimization target. The objective is to minimize the release-relative object completion delay within the bounded candidate plan family:
\begin{equation}
\begin{aligned}
\min_{\chi}\quad
& T^{\mathrm{cmp}}_o(\chi)-t^{\mathrm{rel}}_o \\
\mathrm{s.t.}\quad
& \sum_{k=1}^{K_o} b_{o,k}=B_o,\quad K_o\le K_{\max},\\
& t^{\mathrm{tx}}_{o,k}\ge t^{\mathrm{rel}}_o,\quad
  P_{o,k}\in \Pi_{s_o g_o}\!\left(t^{\mathrm{tx}}_{o,k}\right),\quad \forall k,\\
& \chi \text{ is residual-service consistent.}
\end{aligned}
\end{equation}
Since $t^{\mathrm{rel}}_o$ is fixed for the object, minimizing
$T^{\mathrm{cmp}}_o(\chi)-t^{\mathrm{rel}}_o$ is equivalent to minimizing
the absolute completion time, while making the release-relative delay
explicit. The formulation above leaves chunk sizes and launch times continuous; the restricted exhaustive reference in Sec. VII applies to a finite discretized plan family.

\section{Link-Service Profiles and Single-Link Completion}
\label{sec:s2g_capacity}

This section instantiates the single-link timing primitive used by the finite-object delivery model. The purpose is not to derive a new physical-layer link model, but to map RF/optical link quantities and intermittent visibility into a cumulative-service profile. The resulting service profile is used by the residual evaluator and by the multi-hop earliest-arrival construction of Sec.~VI.




\subsection{Instantaneous Service Rate}

For each directed edge \(e\), we denote by \(C_e(t)\) the
instantaneous communication service rate supplied by the
corresponding RF or optical link. This rate may be instantiated
from a standard RF satellite link budget for space-to-ground
edges~\cite{maral2020satellite} or from an outage-constrained
optical-link model for optical edges~\cite{andrews2005laser,al2001mathematical}.
The present paper does not optimize these physical-layer
parameters; it uses \(C_e(t)\) as a service-profile input to the
finite-object completion and residual-service accounting layer.

\subsection{Visibility and Cumulative Service}
\label{sec:cumulative_service}

A link can serve traffic only when the corresponding contact is active. For a space-to-ground edge, this is captured by the visibility indicator
\begin{equation}
\label{eq:visibility_indicator}
\zeta_e(t)
=
\mathbf{1}\left\{\theta_{\mathrm{el},e}(t)\ge \theta_{\min,e}\right\},
\end{equation}
where \(\theta_{\min,e}\) includes the geometric elevation mask and link-acquisition constraints. For non-space-to-ground edges, \(\zeta_e(t)\) is replaced by the appropriate edge-availability indicator.

The cumulative service delivered by edge \(e\) over \([t,t+\tau]\) is
\begin{equation}
\label{eq:service_process}
S_e(t,\tau)
=
\int_t^{t+\tau} C_e(u)\zeta_e(u)\,du,
\qquad \tau\ge 0 .
\end{equation}
This quantity is measured in bits and combines time-varying rate with intermittent contact availability. It is the unconditioned service profile used for candidate generation and same-model non-striping baseline evaluation. For fixed-plan evaluation, shared service already consumed by earlier transmissions in the candidate plan is represented by the plan-conditioned residual profile \(S^{\mathrm{res}}_{e,j}(t,\tau\mid\chi)\) defined in Sec.~\ref{sec:system_model}.

\subsection{Single-Link Completion}
\label{sec:single_link_completion}

Let
\begin{equation}
\label{eq:propagation_delay}
\delta_e(t)=\frac{d_e(t)}{c}
\end{equation}
be the one-way propagation delay, treated as quasi-static over a transmission epoch. In the empty-service view used for candidate generation and same-model single-path evaluation, the completion delay for \(B\) bits entering edge \(e\) at time \(t\) is
\begin{equation}
\label{eq:edge_completion}
A_e(B;t)
=
\delta_e(t)
+
\inf\left\{
\tau\ge 0:
S_e(t,\tau)\ge B
\right\}.
\end{equation}

For object-level evaluation under a candidate plan \(\chi\), the corresponding residual single-link completion operator for transmission \(j\) is
\begin{equation}
\label{eq:35}
A_{e,j}^{\mathrm{res}}(B;t\mid\chi)
=
\delta_e(t)
+
\inf\left\{
\tau\ge 0:
S_{e,j}^{\mathrm{res}}(t,\tau\mid\chi)\ge B
\right\}.
\end{equation}
When the evaluated transmission is clear from context, we write \(A_e(B;t\mid\chi)\). Equation~\eqref{eq:edge_completion} evaluates the full edge service, while~\eqref{eq:35} evaluates the service state induced by the candidate finite-object plan. The paper does not introduce an independent exogenous backlog variable in the main model; service already consumed within an evaluated plan is represented through \(S^{\mathrm{res}}\).

The corresponding single-link feasibility condition is immediate. A payload of \(B\) bits released on edge \(e\) at time \(t\) can be delivered in full by time \(t+T\) under the unconditioned service profile if
\begin{equation}
\label{eq:single_link_feasibility}
S_e(t,T-\delta_e(t))\ge B,
\qquad
T\ge \delta_e(t).
\end{equation}
Under a candidate plan \(\chi\), the residual-service-consistent condition replaces \(S_e\) by \(S_{e,j}^{\mathrm{res}}(\cdot\mid\chi)\) for the corresponding transmission \(j\). These single-link completion operators are the timing primitives used by the multi-hop earliest-arrival evaluator in Sec.~VI.



\subsection{Event-Grid Representation}
\label{sec:event_grid_representation}

In the numerical implementation, the contact plan is represented
on an event grid augmented with contact-boundary events. The
event-grid approximation and its sensitivity are described with
the multi-hop time-label evaluator in Sec.~VI-D and evaluated
in Sec.~VIII-E.

\section{Multi-Hop Finite-Payload Completion}
\label{sec:isl_routing}

This section lifts the single-link completion operators of Sec.~\ref{sec:s2g_capacity} to finite-payload completion over multi-hop relay paths. The goal is to compute the arrival time of a finite object or chunk over a generated candidate path family, under the same cumulative-service model and FIFO service discipline used in the object-level formulation. Inter-satellite and space-to-ground links enter this construction through their edge service profiles \(C_e(t)\zeta_e(t)\) and propagation delays.

\subsection{Inter-Satellite Service Profiles}
\label{sec:isl_service_profiles}

For optical inter-satellite links, the edge service rate can be instantiated from a standard optical link budget~\cite{kaushal2016optical}. In the numerical study, this physical model is represented by the nominal coded-rate service profile in Table~\ref{tab:sim_settings}. Link availability is captured by the corresponding edge-availability indicator \(\zeta_e(t)\), which encodes the procedural relay topology and any pointing or adjacency constraints used in the controlled contact-plan evaluation. Thus, an inter-satellite link and a space-to-ground link are treated uniformly by the finite-payload timing model: both supply a time-varying cumulative service profile \(S_e(t,\tau)\) and, under a fixed delivery plan, a residual profile \(S^{\mathrm{res}}_{e,j}(t,\tau\mid\chi)\).

\subsection{Finite-Payload Completion on a Fixed Path}
\label{sec:finite_payload}

For a finite payload of \(B\) bits, the relevant timing quantity is the completion time of the whole payload, not asymptotic throughput. Consider a store-and-forward path
\[
P=(e_1,\ldots,e_K).
\]
Under the unconditioned service profile, the per-hop departure and arrival times are defined recursively by
\begin{equation}
\label{eq:hop_recursion}
t_0=t,\qquad
t_k=t_{k-1}+A_{e_k}(B;t_{k-1}),
\quad k=1,\ldots,K,
\end{equation}
where \(A_{e_k}\) is the single-link completion operator in~\eqref{eq:edge_completion}. The corresponding fixed-path delay is
\begin{equation}
\label{eq:path_delay}
A_P(B;t)=t_K-t_0 .
\end{equation}

For object-level delivery under a candidate plan \(\chi\), the path of each chunk is already fixed. The residual path delay is obtained by replacing each unconditioned edge completion with its plan-conditioned residual counterpart:
\begin{equation}
t^{\chi}_0=t,\qquad
t^{\chi}_k=t^{\chi}_{k-1}
+
A^{\mathrm{res}}_{e_k,j_k}
\bigl(B;t^{\chi}_{k-1}\mid\chi\bigr),
\quad k=1,\ldots,K,
\end{equation}
where \(j_k\) is the transmission induced by the evaluated chunk on edge \(e_k\). The resulting residual fixed-path delay is
\begin{equation}
\label{eq:residual_path_delay}
A_P(B;t\mid\chi)=t^{\chi}_K-t^{\chi}_0 .
\end{equation}
This is the path-level quantity used by the finite-object delivery plan in Sec.~\ref{sec:problem_formulation}.

\subsection{Best-Path Finite-Payload Evaluation}
\label{sec:best_path_eval}

When the path has not yet been fixed, for example during candidate generation or same-model single-path baseline evaluation, the best finite-payload delay from node \(m\) to node \(n\) is
\begin{equation}
\label{eq:nodepair_arrival}
A_{m\to n}(B;t)
=
\min_{P\in\Pi^{\mathrm{gen}}_{mn}(t)}
A_P(B;t),
\end{equation}
where \(\Pi^{\mathrm{gen}}_{mn}(t)\) is the generated candidate path family available at release time \(t\). This minimization is not applied again after a delivery plan \(\chi\) has selected a fixed chunk path \(P_{o,k}\). Thus, unconditioned best-path quantities are used for candidate generation and same-model single-path baselines, while \(\chi\)-conditioned fixed-path quantities are used for residual evaluation of a committed plan.

\subsection{Time-Label Evaluation and Event-Grid Approximation}
\label{sec:time_expanded}

The implementation evaluates~\eqref{eq:nodepair_arrival} using a time-label representation over the generated path family. A label consists of a physical node and an arrival time. Waiting advances a label to later event times at the same node. A communication relaxation over edge \(e=(m,n)\) launched from a label time \(t\) creates an arrival label at \(n\) with time \(t+A_e(B;t)\). For fixed-path residual evaluation, the same construction uses the residual edge completion \(A^{\mathrm{res}}_{e,j}(B;t\mid\chi)\).

The label-setting evaluation relies on the non-overtaking property of the single-link completion operator:
\begin{equation}
t'\ge t
\quad\Longrightarrow\quad
t'+A_e(B;t')\ge t+A_e(B;t).
\label{eq:non_overtaking}
\end{equation}
For a fixed service profile and FIFO residual service, delaying a launch cannot create an earlier completion of the same finite payload on the same edge. Under this property, exact insertion of computed arrival labels gives the finite-payload earliest-arrival delay over the generated path family.

The event-grid parameter controls how the time-varying link rate is sampled before these completion operators are evaluated. In the numerical implementation, each event-grid interval is represented by its midpoint rate, and completion within that piecewise-constant interval is computed continuously. The computed arrival time is retained as a continuous label; it is not rounded up to the next event epoch. Consequently, the implemented grid is a midpoint-rate approximation of the continuous service profile, not a conservative upper bound.

The event-grid sensitivity reported in Sec.~VIII-E confirms
that this midpoint-rate approximation is mildly optimistic at
coarse resolutions. We therefore treat event-grid resolution as
a discretization sensitivity rather than a one-sided safety
guarantee.

Together, Secs.~\ref{sec:s2g_capacity} and~\ref{sec:isl_routing} define the communication timing model used by the finite-object delivery problem: a cumulative-service single-link completion relation and a finite-payload multi-hop earliest-arrival evaluator. The structural results that follow concern how these timing relations behave under finite-object, shared-service, and bounded-striping delivery decisions.

\section{Analytical Support and Restricted Reference}
\label{sec:structural_results}

This section gives the analytical support and restricted-reference scope
for the evaluation. We first state a structural observation showing that
streaming-centric path ranking is not generally exact for finite-object
completion. We then give a sufficient condition under which two-way striping
strictly enlarges the deadline-feasible payload region. Finally, we define
the restricted exhaustive reference used for the single-object benchmark and
clarify that it is exact only within its discretized bounded plan family.

\subsection{Streaming-Centric Ranking Failure}

We first record a preliminary characterization that motivates the
cumulative-service evaluator. Finite-payload completion can differ from steady-state throughput; here we express the reversal condition within the present cumulative-service model and use it to support the release-time--payload regime map in Sec. VIII.

Proposition 1 (Streaming-centric ranking failure). There exist two time-respecting paths $P_1$ and $P_2$, a payload size $B$, and a release time $t_0$ such that a streaming-centric criterion ranks $P_1$ above $P_2$, while the finite-payload earliest-arrival metric reverses that ranking. In particular, there exist instances for which
\begin{equation}
\Phi(P_1) > \Phi(P_2),
\end{equation}
yet
\begin{equation}
A_{P_1}(B;t_0) > A_{P_2}(B;t_0),
\end{equation}
where $\Phi(P)$ denotes a streaming-centric path-quality surrogate, such as average rate or bottleneck throughput.

This follows by considering two feasible paths with useful
service onset delays \(w_1>w_2\) measured relative to \(t_0\),
rates \(R_1>R_2\), and propagation delays \(\delta_1,\delta_2\). A rate-based surrogate ranks
\(P_1\) above \(P_2\), but for any payload \(B\) satisfying
\begin{equation}
w_1+\frac{B}{R_1}+\delta_1
>
w_2+\frac{B}{R_2}+\delta_2,
\label{41}
\end{equation}
the finite-payload earliest-arrival order is reversed.

Proposition 1 localizes the ranking reversal within the cumulative-service
model and motivates the release-time--payload regime map in the evaluation.

\subsection{Feasible-Region Expansion Under Two-Way Striping}

The second result formalizes the positive claim that object striping can change delivery behavior in a structural, rather than merely average-performance, sense.

For a fixed release time \(t_0\), a relative deadline budget
\(\Delta\), and an admissible path family \(\mathcal{P}\), let
\(\mathcal{X}^{\mathcal{P}}_K(B;t_0)\) denote the set of delivery plans
\(\chi\) satisfying
\begin{equation}
\begin{aligned}
\mathcal{X}^{\mathcal{P}}_K(B;t_0)
=
\bigl\{\chi:\;&
\sum_{k=1}^{K_o} b_{o,k}=B,\quad K_o\le K,\\
& b_{o,k}>0,\quad k=1,\ldots,K_o,\\
& P_{o,k}\in\mathcal{P},\quad k=1,\ldots,K_o,\\
& t^{\mathrm{tx}}_{o,k}\ge t_0,\quad k=1,\ldots,K_o
\bigr\}.
\end{aligned}
\end{equation}
The candidate-family deadline-feasible payload region under striping
degree at most \(K\) is then
\begin{equation}
\begin{aligned}
\mathcal{F}^{\mathcal{P}}_K(\Delta;t_0)
=
\bigl\{B>0:\;&
\exists\,\chi\in\mathcal{X}^{\mathcal{P}}_K(B;t_0)\\
& \text{s.t. } T_o^{\mathrm{cmp}}(\chi)-t_0\le\Delta
\bigr\}.
\end{aligned}
\end{equation}
This region is defined with respect to the same admissible path family
used by the compared decision layers, rather than an unrestricted global route set.

Proposition 2 (Sufficient deadline service-budget condition for two-way feasible-region expansion). For a release time \(t_0\) and a relative deadline budget
\(\Delta\), define
\begin{equation}
\Delta_1=\Delta-T_o^{\mathrm{reasm}}-\eta_{\mathrm{chunk}},
\qquad
\Delta_2=\Delta-T_o^{\mathrm{reasm}}-2\eta_{\mathrm{chunk}}.
\label{eq:delta_budgets}
\end{equation}
Let
\begin{equation}
Q_P(\tau;t_0)=\sup\{B\ge0:A_P(B;t_0)\le \tau\}
\end{equation}
be the largest payload deliverable by path \(P\) within delay budget
\(\tau\). We refer to \(Q_P(\tau;t_0)\) as the deadline service budget of path \(P\): it is the maximum payload that the cumulative RF/optical service profile along \(P\) can deliver within delay budget \(\tau\). Let \(\mathcal{P}_{\mathrm{single}}\) denote the same admissible path family used for the non-striping single-path comparison. Consider a candidate two-way striping template using two edge-disjoint
paths $P_a, P_b \in \mathcal{P}_{\mathrm{single}}$. Under the per-edge
residual-service model of Sec.~III-D, edge disjointness means that the
two chunks do not consume the same edge service; node-level transmit or
receive coupling at a shared source or destination is not represented in
this per-edge template.
Assume $Q_{P_a}(\Delta_2;t_0)>0$ and $Q_{P_b}(\Delta_2;t_0)>0$.
If \(\Delta_2\ge0\) and the deadline service-budget comparison
\begin{equation}
\label{eq:two_way_service_budget_condition}
 Q_{P_a}(\Delta_2;t_0)+Q_{P_b}(\Delta_2;t_0)>
\max_{P\in\mathcal{P}_{\mathrm{single}}}Q_P(\Delta_1;t_0)
\end{equation}
holds, then there exists a nonempty payload interval
\begin{equation}
I \subseteq
\mathcal{F}^{\mathcal{P}_{\mathrm{single}}}_2(\Delta;t_0)
\setminus
\mathcal{F}^{\mathcal{P}_{\mathrm{single}}}_1(\Delta;t_0).
\end{equation}
The left-hand side of \eqref{eq:two_way_service_budget_condition} is the aggregate pre-deadline service budget provided by two edge-disjoint path profiles after paying the two-chunk overhead, whereas the right-hand side is the largest one-path service budget after paying the one-chunk overhead. Thus, the condition identifies when complementary communication-service opportunities, rather than path
multiplicity alone, are sufficient to enlarge the object-level deadline-feasible payload region.
\begin{IEEEproof}
Let
\begin{equation}
B_{\mathrm{single}}
=
\max_{P\in\mathcal{P}_{\mathrm{single}}}Q_P(\Delta_1;t_0).
\end{equation}
By monotonicity of \(A_P(\cdot;t_0)\) in payload and the non-overtaking
property in ~\eqref{eq:non_overtaking}, the single-path candidate-family feasible region is the
down-closed interval \((0,B_{\mathrm{single}}]\). The strict inequality gives a nonempty interval
\begin{equation}
\left(
B_{\mathrm{single}},
Q_{P_a}(\Delta_2;t_0)+Q_{P_b}(\Delta_2;t_0)
\right].
\end{equation}
Choose \(B\) in this interval and split it into positive
\(b_a+b_b=B\) with
\(b_a\le Q_{P_a}(\Delta_2;t_0)\) and
\(b_b\le Q_{P_b}(\Delta_2;t_0)\).  Since \(P_a\) and \(P_b\) are
edge-disjoint under the per-edge residual-service model, the
\(\chi\)-conditioned residual service on each selected path equals the
corresponding unconditioned single-path service used to define
\(Q_{P_a}\) and \(Q_{P_b}\) for this two-way template. Thus both chunks
arrive within the two-chunk arrival budget \(\Delta_2\), and the two
chunk overheads plus reassembly fit within the relative deadline budget
\(\Delta\). Since \(B>B_{\mathrm{single}}\), no path in
\(\mathcal{P}_{\mathrm{single}}\) can deliver payload \(B\) within the
single-chunk budget \(\Delta_1\). Hence \(B\) belongs to the two-way
candidate-family feasible region but not to the corresponding single-path
candidate-family feasible region.
\end{IEEEproof}

Proposition 2 applies to edge-disjoint complementary contacts under the
per-edge residual-service model. Since \(P_a\) and \(P_b\) are edge-disjoint,
the \(\chi\)-conditioned residual service on each selected path equals the
corresponding unconditioned fixed-path service. Proposition~2 therefore
characterizes a complementary-contact striping gain rather than an empirical
activation of residual accounting. The residual-service layer becomes
non-trivial under shared-edge or cross-object contention, which is evaluated
separately in Sec.~VIII.

Node-level transmit/receive coupling at a shared source or destination,
such as a single ground-station receiver, requires an additional resource
constraint; hence the gain is per-edge rather than receiver-scheduled.

For numerical evaluation, we use the following deadline service-budget score
corresponding to Proposition 2. With \(\Delta_1\) and \(\Delta_2\) as
defined in \eqref{eq:delta_budgets}, let
\(\mathcal{P}_{\mathrm{single}}\) denote the single-path candidate family
used by the non-striping baseline, and let \((P_a,P_b)\) be a candidate
two-way striping template. A sufficient deadline service-budget score for
candidate-family two-way frontier enlargement is
\begin{equation}
\begin{aligned}
S_{\mathrm{strip}}(\Delta;t_0,P_a,P_b)
&=
Q_{P_a}(\Delta_2;t_0)+Q_{P_b}(\Delta_2;t_0) \\
&\quad -
\max_{P\in\mathcal{P}_{\mathrm{single}}} Q_P(\Delta_1;t_0).
\label{eq47}
\end{aligned}
\end{equation}

When \(S_{\mathrm{strip}}>0\), the two paths jointly provide more
pre-deadline payload budget after paying the additional chunk overhead
than the best single-path delivery in \(\mathcal{P}_{\mathrm{single}}\)
can provide with one chunk. Under the edge-disjoint, residual-service-consistent template considered here, this gives a sufficient condition for a nonempty payload interval that is feasible under the two-way candidate family but infeasible under the corresponding single-path candidate family. Thus, \(S_{\mathrm{strip}}>0\) is used as a sufficient screening score for the edge-disjoint
two-way templates evaluated in Sec.~VIII. It is a sufficient diagnostic for complementary-contact gain, not a necessary condition for every possible striping benefit.

\subsection{Restricted Exhaustive Reference}
\label{subsec:restricted_reference}

We next clarify the restricted setting under which the exhaustive reference
is exact. The reference is restricted to a discretized single-object bounded
plan family: the striping degree is bounded by the two-way plan family, the
candidate path set is finite, the chunk partitions and launch-time grid are
finite, and each chunk follows a fixed path with non-preemptive transmission.
The FIFO service discipline and deterministic tie-breaking rule assign a
unique completion value to every feasible plan under residual-service-consistent
accounting. Exhaustive enumeration over this finite plan family therefore
returns the exact optimum for the restricted benchmark problem.

The scope of this restricted exhaustive reference is the discretized \(K=2\) plan family; it is not a global optimum or an upper bound on every implemented scheduler. It is exact only within the discretized \(K=2\) plan family described above. The scalable scheduler used in the system experiments additionally includes a quantized
greedy fallback outside this restricted family; as a result, signed gaps to
the restricted reference can be negative on some instances. For comparability
with the original benchmark summary, the main gap table reports the
absolute-deviation convention, while the signed behavior is reported in the
tail diagnosis.

We verified the implementation of this restricted reference against an
independent un-pruned full enumeration on small discretized plan sets,
including a prune-firing relaxed-deadline case. On every tested instance,
the implementation returned the same lexicographic optimum as un-pruned
enumeration under the objective
\[
\text{on-time} \succ \text{delivered} \succ -\text{lateness} \succ -\text{makespan}.
\]
The branch-and-bound rule prunes only branches that are strictly suboptimal
in the primary on-time objective and therefore preserves delivered-volume,
lateness, and makespan tie-breakers under a primary-objective tie. Continuous
launch times, larger restricted striping degrees, node-level receiver
coupling, and full multi-object scheduling are outside this finite enumeration.

\section{Numerical Evaluation}
\label{sec:experiments}

This section evaluates the analytical claims and service-layer behavior under a shared controlled communication setting. Unless stated otherwise, the compared methods use the same procedural Walker-like contact model, RF/optical service profiles, finite-payload timing model, FIFO service discipline, and candidate path family. The evaluation is a controlled same-model characterization of the delivery layer rather than a protocol-level competition with full CGR/SABR implementations or a deployment trace.

\begin{table}[t]
\centering
\caption{Compared decision layers.}
\label{tab:methods}
\footnotesize
\setlength{\tabcolsep}{3pt}
\renewcommand{\arraystretch}{1.08}
\begin{tabularx}{\columnwidth}{@{}p{0.28\columnwidth}cX@{}}
\toprule
Method & Striping & Decision layer and role \\
\midrule
Streaming surrogate
& no
& Average-rate or bottleneck-style path preference; ranking-failure baseline. \\

Single-path baseline
& no
& Best finite-payload earliest-arrival path; same-model non-striping baseline. \\

Bounded two-way striping
& yes
& Bounded two-way plan family evaluated with the same finite-payload timing model. \\

Restricted exhaustive reference
& bounded
& Exhaustive enumeration over a discretized \(K=2\) plan family; reference for the single-object benchmark. \\
\bottomrule
\end{tabularx}
\end{table}

\begin{table}[t]
\centering
\caption{Main controlled-evaluation parameters.}
\label{tab:sim_settings}
\scriptsize
\setlength{\tabcolsep}{2pt}
\renewcommand{\arraystretch}{1.06}
\begin{tabularx}{\columnwidth}{@{}p{0.27\columnwidth}p{0.35\columnwidth}X@{}}
\toprule
Item & Baseline value & Variation / usage \\
\midrule
Main benchmark & one object, one source, one GS & background objects only in residual-accounting diagnostics \\
Orbit & 550 km, \(53^\circ\) inclination & fixed \\
Contact model & procedural Walker-like topology & controlled contact-plan evaluation \\
Baseline constellation & 4 planes, 5 satellites/plane & scalability sweep: 20--180 satellites \\
ISL topology & up to 4 optical ISLs/sat. & intra-/inter-plane neighbours \\
Ground station & \(45^\circ\) lat., \(0^\circ\) lon. & fixed unless swept \\
Minimum elevation & \(15^\circ\) & communication sensitivity sweep \\
RF downlink & Ka-band, 26.5 GHz, 0.10 GHz BW & communication sensitivity sweep \\
Optical ISL & 1.0 Gbps nominal coded rate & controlled contact-plan tests \\
Event grid & 20 s, horizon \(1.6T_{\mathrm{orb}}\) & event-grid sensitivity: 5--60 s \\
Service discipline & per-edge FIFO residual service & residual evaluations \\
Restricted reference & discretized \(K=2\) plan family & exact only within this family \\
Scalable scheduler & bounded search with greedy fallback & fallback lies outside restricted reference \\
Chunk overhead & 0 s unless swept & overhead-sensitivity sweep \\
Chunk quantum & 60 Mbit & tail decomposition: 60/30/20 Mbit \\
Candidate paths & up to 6 paths/chunk & tail decomposition to 12 paths / 66 pairs \\
Relay-hop limit & up to 3 relay hops & controlled benchmark \\
Frontier grid & 1.2--7.2 Gbit, 6 payload values & 9 deadline-slack values \\
Release screening & 10 release samples & multi-release summary \\
Reference benchmark & 32 reference-active instances & 18 releases, 8 payload probes \\
\bottomrule
\end{tabularx}
\end{table}

\subsection{Shared-Contact Residual Accounting}
\label{subsec:a2_contention}

We first evaluate the residual layer where it is non-trivial: same-object chunks or controlled competing objects share contact service. This diagnostic uses the same service profiles and FIFO discipline as the rest of the evaluation and isolates shared-service accounting for committed plans.

\begin{figure*}[t]
\centering
\includegraphics[width=\linewidth]{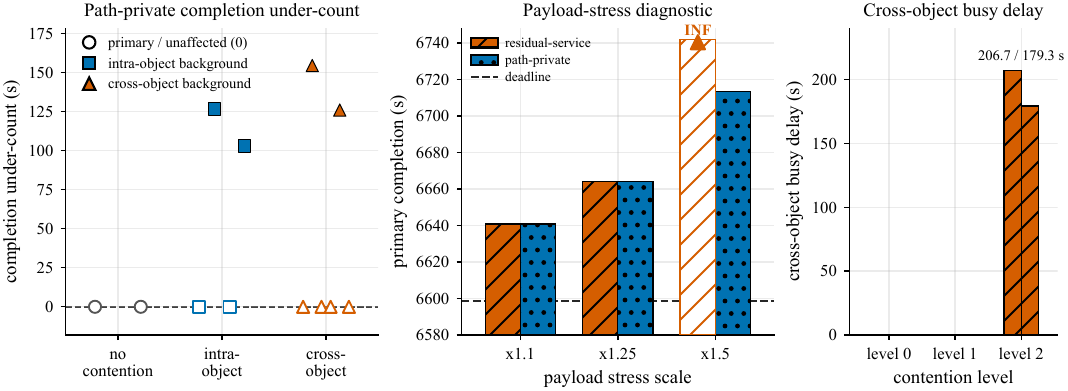}
\caption{Residual-service accounting under shared-contact contention.}
\label{fig:a2_contention}
\vspace{-4pt}
\end{figure*}

Figure~\ref{fig:a2_contention} compares path-private and residual evaluation on the same committed plans. In Fig.~\ref{fig:a2_contention}, \(0\) under-count means that path-private and residual evaluation agree on the same committed plan. The payload-stress panel uses a truncated y-axis at 6580 s; its dashed line marks the deadline. At the higher contention level, the residual busy pointer delays the later transmission by 179--207 s, and realized residual windows do not overlap. A path-private evaluation under-counts completion by up to 154.3 s when transmissions share a contact. The under-count has two sources: intra-object serialization, when many chunks of one object reuse the same edge, and cross-object serialization, when controlled background traffic contends for a shared direct downlink. The highest-priority object is not pushed in this diagnostic, so the under-count falls on lower-priority background traffic.

Under a payload-stress diagnostic, the residual evaluator finds no residual-service completion for a fixed plan while the path-private view reports a finite completion that is still past the deadline. The stress case exposes fixed-plan accounting failure rather than deadline success. The busy-delay panel confirms the same.

The ETO-style sequential-booking baseline and the joint residual scheduler select different background placements under contention, but their primary-object completion is identical in this priority-ordered setting. We therefore use this experiment to demonstrate accounting correctness and decision-layer separation under the tested priority ordering.

\subsection{Streaming-Centric Ranking Reversal}
\label{subsec:e1}

\begin{figure*}[t]
\centering
\includegraphics[width=0.9\linewidth]{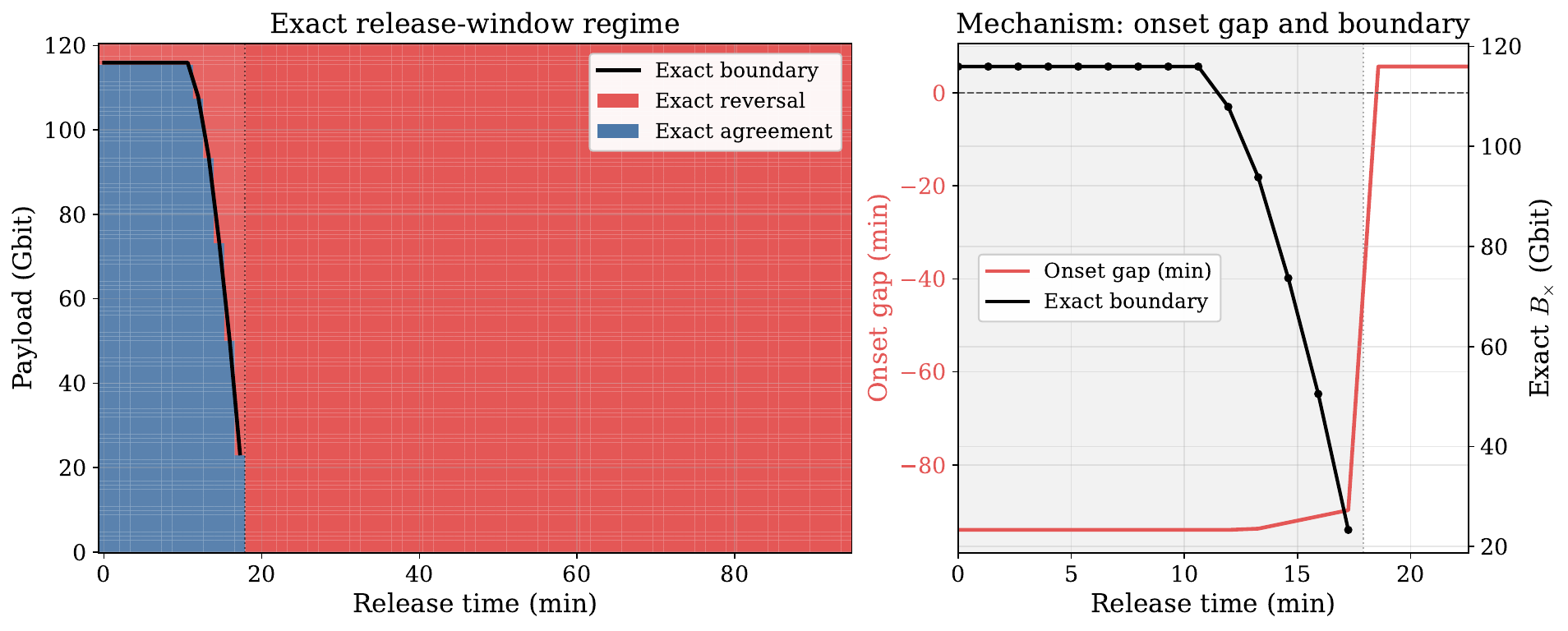}
\caption{Release-time--payload regime in which streaming-centric path ranking reverses finite-payload earliest arrival.}
\label{fig:e1_reversal_region}
\vspace{-4pt}
\end{figure*}

We next use Proposition~1 to illustrate why finite-payload completion cannot be reduced to a streaming-centric path ranking. Figure~\ref{fig:e1_reversal_region} maps the release-time--payload regimes for a representative path pair. Red cells mark cases where the streaming-preferred path arrives later than the alternative finite-payload path; blue cells mark agreement. The reversal occupies a nonempty region of the tested plane, and the crossover boundary appears for 14 of the 72 release-grid points used in Fig.~\ref{fig:e1_reversal_region}, from 0.00 to 17.24 min.

This result supports the use of the cumulative-service finite-payload evaluator. It is separate from the residual-accounting diagnostic above: ranking reversal concerns path choice under intermittent service, whereas residual accounting concerns shared-service consistency for a committed plan.

\subsection{Complementary-Contact Two-Way Gain}
\label{subsec:e2}

We next evaluate Proposition~2 in complementary-contact regimes. This part of the evaluation concerns edge-disjoint timing gain: in this regime, residual service reduces to unconditioned fixed-path service, and the gain comes from complementary contact timing rather than from residual-accounting activation.

\begin{figure*}[t]
\centering
\includegraphics[width=\linewidth]{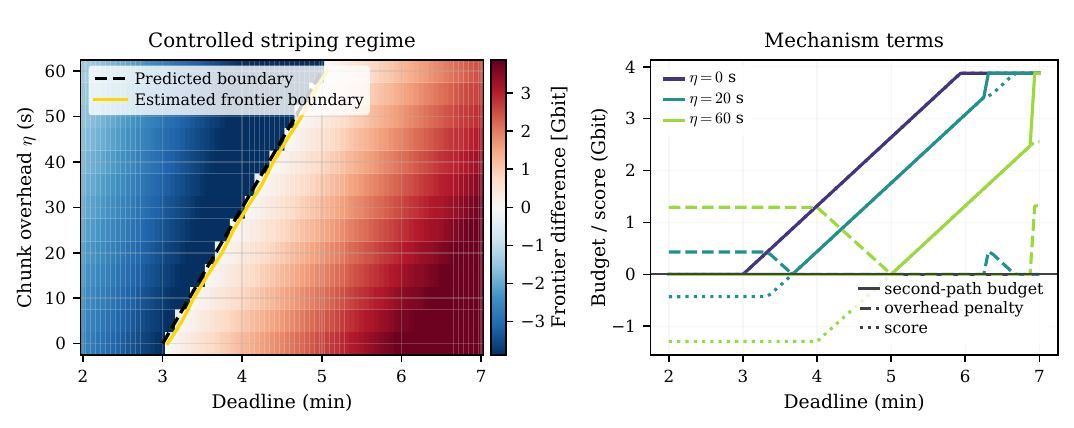}
\caption{Controlled deadline service-budget criterion for identifying when two-way striping enlarges the feasible frontier.}
\label{fig:e2_controlled_geometry}
\vspace{-4pt}
\end{figure*}

Figure~\ref{fig:e2_controlled_geometry} first operationalizes Proposition~2 in the controlled complementary-contact construction. The left panel reports the signed frontier difference of a candidate two-way striping template relative to the same-model non-striping baseline over deadline and equivalent chunk overhead. The dashed curve is the predicted zero boundary obtained from \(S_{\mathrm{strip}}=0\), while the yellow curve is the bisection-estimated frontier boundary at which the measured template-level frontier difference changes sign.

The two boundaries are close over the tested operating region, indicating that the deadline service-budget score captures the onset of useful striping in this controlled construction. The negative region is a template-level loss: under those deadlines or overheads, the evaluated two-way template does not compensate for the additional chunk overhead. It does not imply that the optimized \(K_o\le 2\) feasible envelope is worse than the \(K_o=1\) envelope, since the latter can be recovered by single-path fallback. The positive region corresponds to the regime in which the second-path pre-deadline budget exceeds the overhead-induced loss relative to the best single-path delivery.

The controlled construction confirms the service-budget interpretation of Proposition~2. Strict expansion is observed, with an expansion fraction of 0.1667 and a maximum payload lift of 3.840 Gbit. The maximum deadline saving is 0, so in this controlled construction the enlargement is expressed primarily along the payload dimension.

\begin{figure}[t]
\centering
\includegraphics[width=\linewidth]{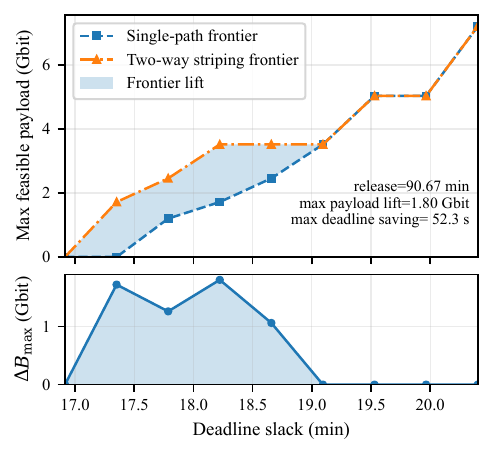}
\caption{Dense feasible-region frontier for the selected release in the complementary-contact source context of the procedural Walker-like contact model.}
\label{fig:e2_real_geometry}
\vspace{-6pt}
\end{figure}

We then evaluate the selected complementary-contact source context in the procedural Walker-like contact model. Over a release distribution in this context, 9 of 10 sampled releases yield positive gain and 8 of 10 strictly enlarge the feasible region, with a grid-quantized payload lift of about 1.8\,Gbit and up to about 42\,s deadline saving; the selected release in Fig.~\ref{fig:e2_real_geometry} is representative of this positive-gain context rather than an outlier. A second source context yields no gain in any sampled release, confirming that the effect is context-dependent rather than universal.


Figure~\ref{fig:e2_real_geometry} gives the denser feasible-region frontier for the selected release. In this dense frontier calculation, the expansion fraction is 0.1296, the maximum payload lift is 1.799\,Gbit, and the maximum deadline saving is 52.282\,s. The screening summary is coarser than the dense selected-release frontier, which is why the reported deadline-saving values differ.

The controlled and procedural Walker-like results support the complementary-contact behavior captured by Proposition~2. Two-way striping creates strict expansion in complementary-contact regimes, while the gain remains release-, payload-, and context-dependent.

\subsection{Gap to the Restricted Exhaustive Reference}
\label{subsec:e3}

We next compare the single-path baseline and bounded two-way striping against the restricted exhaustive reference. The reference is exact only within the discretized \(K=2\) plan family described in Sec.~VII; it is not a global optimum or an upper bound on the implemented scheduler. The comparison uses 32 reference-active single-object benchmark instances selected from 18 release samples and 8 payload probes. Selection requires that the restricted reference use a nontrivial multi-path or multi-chunk plan and that the single-path baseline have a nonzero gap; it does not condition on the bounded two-way scheduler's gap.

\begin{figure}[t]
\centering
\includegraphics[width=\linewidth]{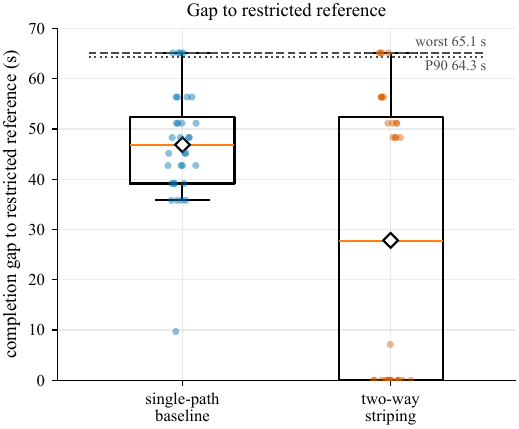}
\caption{Completion gap to the restricted exhaustive reference over 32 reference-active single-object benchmark instances.}
\label{fig:e3_benchmark}
\vspace{-4pt}
\end{figure}

Figure~\ref{fig:e3_benchmark} reports the absolute completion gap to the restricted reference. Bounded two-way striping reduces the mean and median gaps from 46.87 s and 46.74 s to 27.85 s and 27.73 s, respectively, about a 40\% reduction relative to the same-model single-path baseline. The P90 and worst-case gaps remain unchanged at 64.27 s and 65.15 s.

The unchanged upper tail is not a search-resolution artifact. Relaxing the candidate-path budget to 66 pairs, chunk granularity to 20 Mbit, and launch slots to 16 recovers none of the P90 or worst-case gap, while runtime rises. A plan-disagreement diagnostic shows that on all four P90/worst-case instances the scheduler commits a single-path plan whereas the restricted reference uses two-way striping. Thus, the residual tail is associated with a single-path-versus-two-way plan-selection disagreement rather than insufficient path, split, or launch-grid resolution.

\begin{table}[t]
\centering
\caption{Gap to the restricted exhaustive reference and runtime summary.}
\label{tab:e3_benchmark_summary}
\scriptsize
\setlength{\tabcolsep}{3.5pt}
\begin{tabular}{lccccccc}
\toprule
Method & Mean & Med. & P90 & Max & Mean & Med. & P90 \\
& gap (s) & gap (s) & gap (s) & gap (s) & rt. (s) & rt. (s) & rt. (s) \\
\midrule
Restricted reference
& 0.000 & 0.000 & 0.000 & 0.000
& 0.387 & 0.361 & 0.640 \\

Single-path\\baseline
& 46.874 & 46.738 & 64.266 & 65.145
& 0.0166 & 0.0165 & 0.0171 \\

Two-way\\striping
& 27.845 & 27.725 & 64.266 & 65.145
& 0.268 & 0.199 & 0.672 \\
\bottomrule
\end{tabular}
\end{table}
\vspace{-6pt}

Absolute gaps are reported for comparability; signed gaps can be negative because the scalable scheduler includes a greedy fallback outside the restricted \(K=2\) reference family.

\subsection{Scalability and Event-Grid Sensitivity}
\label{subsec:scalability}

Finally, we evaluate the delivery layer as a controlled system component. The scalability sweep uses a feasibility-aware workload builder over the procedural Walker-like contact model. The sweep characterizes computational tractability and implementation boundaries under the controlled contact-plan model.

\begin{figure}[t]
\centering
\includegraphics[width=\linewidth]{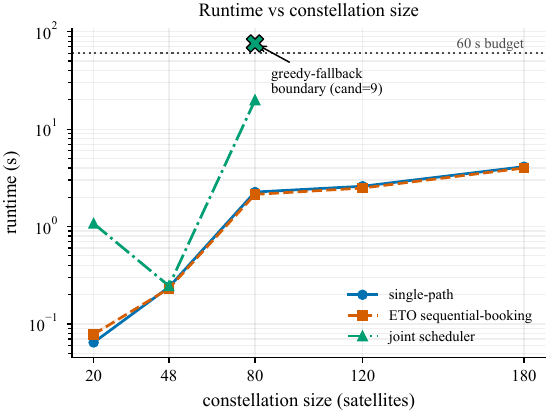}
\caption{Runtime scalability of the delivery layer under the procedural Walker-like contact model.}
\label{fig:ea_runtime}
\vspace{-4pt}
\end{figure}

Figure~\ref{fig:ea_runtime} reports policy runtime as constellation size and candidate-path budget increase. The single-path and ETO-style sequential-booking arms remain tractable across the 20--180 satellite ladder, with mean runtime increasing from about 0.06 s to about 4.1 s. The joint scheduler is tractable at smaller sizes and lower candidate budgets, but at 80 satellites with candidate budget 9 its quantized greedy fallback triggers many chunks and exceeds the 60 s per-run budget in the P90 case. We therefore mark this setting as a greedy-fallback runtime boundary rather than extrapolating the joint scheduler to 120 or 180 satellites.

At the 80-satellite setting, increasing the candidate-path budget exposes the same faster path to all three arms. The completion improvement is therefore a candidate-path-budget effect, not a joint-scheduler performance gain. Under the feasibility-aware workload, primary completion remains finite across the evaluated 20--180-satellite ladder. The cross-object contention signal is geometry-dependent, present at 20 and 80 satellites and absent for the evaluated arms at 48, 120, and 180 satellites; it is not monotone in constellation size.

The event-grid sensitivity check confirms the discretization behavior described in Sec.~VI. Relative to a \(5\) s grid, the baseline \(20\) s grid gives completion times about \(1\) s smaller on average for the tested objects, i.e., a small optimistic bias rather than a conservative delay. Coarser \(40\)--\(60\) s grids remain within about \(15\) s worst observed completion-time bias, less than \(0.25\%\) of the corresponding completion times, while reducing per-evaluation cost by about \(3.5\times\). Thus, event-grid resolution is treated as a small discretization sensitivity, not a one-sided safety guarantee.

The communication-side sensitivity sweeps in the original controlled setting show that RF bandwidth, elevation mask, and chunk overhead affect the returned two-way gain through the cumulative-service budget \(S_{\mathrm{strip}}\). Positive returned-plan scores remain empirically associated with useful two-way striping, consistent with the sufficient condition of Proposition~2. We omit the detailed sensitivity figure for space; the retained runtime and event-grid results are used as a controlled characterization of the service layer rather than as deployment validation.

\section{Conclusion}
\label{sec:conclusion}

This paper studied residual-service-aware finite-object delivery over intermittent RF/optical LEO contact plans. The proposed service layer decides whether a complete deadline-bound object reaches the ground after candidate paths have been generated. The main result is that path-private evaluation can under-count completion when same-object chunks or controlled competing objects share a contact, whereas residual evaluation keeps the committed plan service-consistent. This correctness role is distinct from the edge-disjoint complementary-contact gain, under which two-way striping can enlarge the deadline-feasible payload region. Against the restricted exhaustive reference, bounded two-way search improves mean and median completion gaps over the same-model single-path baseline, while the P90 and worst-case gaps remain unchanged; the tail diagnosis associates these cases with single-path-versus-two-way plan selection rather than search-grid resolution. Controlled scalability experiments show tractable light-arm runtime over a 20--180-satellite procedural contact model and expose a joint-scheduler greedy-fallback boundary. Future work includes multi-object scheduling, node-level coupling, higher striping degrees, and SGP4/TLE-based CGR/SABR integration.

\bibliographystyle{ieeetr}
\bibliography{references}

\end{document}